\documentstyle[12pt]{article}
\newcommand{\be}{\begin{equation}}
\newcommand{\ee}{\end{equation}}
\newcommand{\bea}{\begin{eqnarray}}
\newcommand{\eea}{\end{eqnarray}}
\newcommand{\bi}{\begin{itemize}}
\newcommand{\ei}{\end{itemize}}
\newcommand{\bc}{\begin{center}}
\newcommand{\ec}{\end{center}}
\newcommand{\bfl}{\begin{flushleft}}
\newcommand{\efl}{\end{flushleft}}
\newcommand{\bfr}{\begin{flushright}}
\newcommand{\efr}{\end{flushright}}
\newcommand{\f}{\frac}

\newcommand{\equ}[1]{(\ref{#1})}

\def\6{\partial} \def\a{\alpha} 
 \def\d{\delta} 
\def\e{\epsilon}
  \def\th{\theta}
  \def\l{\lambda}
\def\m{\mu}  \def\x{\xi} \def\P{\Pi}

\def\o{\omega}  \def\D{\Delta}
 \def\L{\Lambda} 
\def\PH{\Phi}

\def\non{\nonumber\\}
\def\={\!\!\!&=&\!\!\!}
\def\+{\!\!\!&&\!\!\!+~}
\def\-{\!\!\!&&\!\!\!-~}

\renewcommand{\AA}{{\cal A}}

\newcommand{\CC}{{\cal C}}

\newcommand{\MM}{{\cal M}}

\newcommand{\journal}[4]{{\em #1~}#2\,(19#3)\,#4;}

\newcommand{\ijmp}{\journal {Int. J. Mod. Phys.}}

\newcommand{\pr}{\journal {Phys. Rev.}}

\newcommand{\jmp}{\journal {J. Math. Phys.}}

\newcommand{\cmp}{\journal {Comm. Math. Phys.}}
\newcommand{\cqg}{\journal {Class. Quantum Grav.}}

\newcommand{\np}{\journal {Nucl. Phys.}}
\newcommand{\pl}{\journal {Phys. Lett.}}

\newcommand{\annp}{\journal {Ann. Phys. (N.Y.)}}

\begin{document}
\title{BRST COHOMOLOGY OF THE SUPERSTRING IN
SUPER-BELTRAMI PARAMETRIZATION}
\author{{\em Liviu T\u{a}taru and Ion V. Vancea,}  \\
Department of Theoretical
Physics}
\date{\today}

\maketitle
\begin{abstract}
A method for the calculation of the BRST cohomology, recently
developed
for 2D gravity theory and the bosonic string in
Beltrami parametrization, is generalized to the superstring  theories
quantized in super-Beltrami parametrization.
\end{abstract}
\setcounter{page}{0}
\thispagestyle{empty}
\newpage

\section{Introduction}
The BRST method for quantization of general theories is the most
important method of quantization and it has been applied to all
gauge theories
\cite{ht}. In this framework the search for the invariant
Lagrangian,
the anomalies and the Schwinger terms corresponding to a given set
of field transformations can be done in a purely algebraic way by
solving the BRST consistency condition in the space of the
integrated
local field polynomials ~\cite{wz1,s,wss,bdk}. This amounts to
study
the nontrivial solutions of the equation
\be
s\AA=0
\label{cons}
\ee
with $s$ the nilpotent BRST differential and $\AA$ and integrated
local functional $\AA=\int d^2 x f$. The condition ~(\ref{cons})
translates into the local descent equations \cite{s}:
\bea
s\omega_2+d\omega_1=0, \hspace{2cm} s\o_1+d\o_0=0 \non
s\o_0=0
\label{des}
\eea
where $\o_2$ is a 2-form with $\AA=\int\o_2$ and $\o_1$, $\o_0$ are
{\em local} 1- and 0-forms. It is well known \cite{wss,bdk,bss}
that the descent equations terminate in the bosonic string or
the superstring in Beltrami or super-Beltrami parametrization,
always with a nontrivial 0-form $\o_2$ and that their "integration"
is trivial
\be
\o_1=\d\o_0 \hspace{2cm} \o_2=\f1{2}\d^2\o_0  \label{delta}
\ee
where the operator $\d$ was introduced by Sorella \cite{s} and it
allows to express the exterior derivative $d$ as a BRST commutator
\be
d=-[s,\d].\label{sorella}
\ee
Thus it is sufficient to determine the general solution of
\be
s\o_0=0
\label{basic}
\ee
in the space of local functions of the fields and their derivatives,
 i.e.
to calculate the BRST cohomology group H(s).

In this paper we shall investigate the structure of H(s) for the
superstring
theory in the super-Beltrami parametrization. The Beltrami superfield
parametrization was introduced for the first time in physics by
Gates and Nishino and independently by Rocek and al.\cite{gat}. Further
studies of heterotic strings in Beltrami superfield formulation as well
as 2D supergravity were made in \cite{br,gir} and more recently in \cite{gr}.
We shall also make use of
the results for the superfield as well as for the component field
formalism obtained in \cite{bbg1,bbg2,dg,g,bss}. The
investigation of~(\ref{basic}) is considerable simplified by
introducing
an appropriate new basis of variables substituting the field and
their
derivatives. The way in which this new basis is chosen is a crucial
step in the calculation.

The paper is organized as follows. In Sec.2 we briefly recall the
BRST symmetry within the superstring in the super-Beltrami
parametrization. In Sec.3 we split the algebra of fields and their
derivatives $\AA$ in the contractive part $\CC$ and the minimal
part $\MM$ by using the famous Sullivan Theorem \cite{sull}. In
Sec.4 we introduce a new basis and show how we can calculate the
elements of H(s).At the end we shall give some explicit examples
including the classical action $S_0$ and the anomalies.

\section{BRST symmetry for the superstring}
\setcounter{equation}{0}
Let us start by introducing the setup for the superstring in the
super-Beltrami parametrization. We shall work on a Riemann surface,
i.e.
a real, smooth 2-dimensional manifold with a positive definite
metric and the
local coordinates $(x,y)$ or complex  coordinates $(z=x+iy~,
~\bar{z}=x-iy)$.
 The classical fields are
$\{X,\l,\bar{\l}, F, \m, \bar{\m}, \a, \bar{\a}\}$
where $X=(X^\m)$ are the string coordinates, $\l,\bar{\l}$ are
their fermionic superpartners and $F$ are auxiliary fields,
$\m$ stands for the bosonic Beltrami differential and $\a$, called
"Beltramino", is its fermionic superpartner. The doublets $(\m,\a)$
and $(\bar{\m}, \bar{\a})$ characterize the super-Beltrami
parametrization in the 2d-superspace. These  variables describe
 the full (1,1) superstring
theory. According to Delduc and Gieres \cite{dg}, the (1,0) version
can be obtained by truncation $\bar{\a}=\bar{\l}=F=0$, i.e. in this
case the supersymmetry is present only in one sector, the other
being just the bosonic string.

We shall suppose that the theory is invariant under 2d
superdiffeomorphism
which is expressed by the following BRST transformations \cite{dg}:
\be
s\PH=c\PH_{1,0}+\bar{c}\PH_{0,1}+\f1{2}\e\PH_{\e}+
\f1{2}\bar{\e}\PH_{\bar{\e}}+
a_{\PH}(\6 c)\PH+b_{\PH}(\bar{\6}\bar{c})\PH\label{brst1}
\ee
with $\PH=\{X,\l,\bar{\l},F\}$ and
\be
\PH_{1,0}=D_z\PH \hspace{1cm},\hspace{1cm} \PH_{0,1}=D_{\bar{z}}\PH
\ee
The supercovariant derivatives $D_z\PH$ and $D_{\bar{z}}\PH$
for $\PH=\{X, \l, \bar{\l}, F\}$   are
defined as in \cite{dg}:
\bea
D_z X=\f1{1-\m\bar{\m}}[ (\6-\bar{\m}\bar{\6})X+\f{1}{2}\bar{\m}
\a\l-\f1{2}\bar{\a}\bar{\l}] \\
D_z\l=\f{1}{1-\m\bar{\m}} [(\6-\bar{\m}\bar{\6}+
\f{1}{2}\bar{\m}\6\m)\l
-\f1{2}\bar{\m}\a D_z X+\f{i}{2}\bar{\a} F]\\
D_{\bar{z}}\l=\f{1}{1-\m\bar{\m}} [(\bar{\6}-\m\6-\f{1}{2}\6\m\l
-\f1{2}\a (D_z X) -\f{i}{2}\m\bar{\a} F]\\
D_z F=\f{1}{1-\m\bar{\m}}[(\6-\bar{\m}\bar{\6}-
\f1{2}\bar{\6}\bar{\m}+
\f1{2}\bar{\m}\6\m)F-\f{i}{2}\bar{\a} (D_{\bar{z}}\l)-
\f{i}{2}\bar{\m}\a
(D_{z}\bar{\l})]
\label{covariant}
\eea
and their complex conjugates, which are obtained by putting bars
everywhere and replacing $'i'$ by $'-i'$. In \equ{covariant} we
have used the notations $\6=\6_z=\6/\6 z$
and $\bar{\6}=\6_{\bar{z}}=\6/\6\bar{z}$.
The fields $\PH_\e$ and $\PH_{\bar{\e}}$
are
\bea
X_\e &=& \l                 \hspace{1.5cm}  ,\hspace{1cm}
 X_{\bar{\e}}~~=~~\bar{\l} \\  \label{e1}
\l_{\e} &=& X^{1,0}         \hspace{1cm},\hspace{1.07cm}
 {\l}_{\bar{\e}}~~=~~-iF \\  \label{e2}
\bar{\l}_{\e} &=& iF        \hspace{1.3cm},\hspace{1.1cm}
 \bar{\l}_{\bar{\e}}~~=~~X^{0,1} \\  \label{e3}
F_\e &=& -i{\bar{\l}}^{1,0} \hspace{0.7cm},\hspace{1.07cm}
 F_{\bar{\e}}~~=~~i\l^{1,0}  \label{e4}
\eea
The numerical coefficients $a_\PH$ and $b_\PH$, which will
 play a crucial
role in the following, have the values:
\bea
a_X&=&b_X=0 \hspace{1cm}, \hspace{1cm} a_F=\f1{2},
\/b_F=\f1{2}\non
a_\l&=&\f{1}{2},\/b_\l=0 \hspace{1cm}, \hspace{1cm}
a_{\bar{\l}}=0,\/
b_{\bar{\l}}=\f1{2}.
\eea
The pairs $(a_{\PH}, b_{\PH})$ is, by definition,
{\em the total weight}
of the field $\PH$. The BRST transformations of the
super-Beltrami
differentials $(\m\/,\/\a)$ are given by:
\bea
s\m&=&\bar{\6}c-\m\6c+(\6\m)c+\f{1}{2}\a\e \label{brst2}\\
s\a&=&\bar{\6}\e-\m\6\e+\f{1}{2}(\6\m)\e+c(\6\a)+\f{1}{2}\a\6c.
\label{brst3}
\eea
and their c.c. Finally the BRST transformations of the ghosts
$(c\/,\/\bar{c})$ and $(\e\/,\/\bar{\e})$ are:
\bea
sc&=&c\6c-\f1{4}\e \e \label{brst4} \\
s\e&=&c\6\e-\f1{2}\e\6c\label{brst5}
\eea
and their c.c. so that
\be
s^2=0.
\label{patr}
\ee
The fermionic ghosts $(c\/,\/\bar{c})$ and their {\em bosonic}
superpartners $(\e\/,\/\bar{\e})$ are related to the usual
superdiffeomorphism ghosts $(\x\/,\/\bar{x})$ and $(\x^\th\/,
\/\bar{\x}^\th)$
by ~\cite{dg}
\be
c=\x+\m\bar{\x} \hspace{1cm},\hspace{1cm} \e=\x +{\bar\x}\a ,
\ee
and their c.c.

For the superstring in the super-Beltrami parametrization
the operator $\d$,
introduced by Sorella \cite{s,wss,bss}, which satisfies
eq.\equ{sorella}
is defined by
\bea
\label{delta1}
\d c   & = & dz+\m d\bar{z} \hspace{1cm}  ,
\hspace{1cm} \d\bar{c}~~=~~d\bar{z}+\bar{\m} d z \non
\d\e   & = & -\a d\bar{z}   \hspace{1.6cm},
\hspace{1cm} \d\bar{\e}~~=~~-\bar{\a}d z\non
\d\Phi & = & 0 \hspace{2.3cm}\mbox{for}\hspace{0.9cm}
\Phi~~=~~\{X,\l,\bar{\l},F,\m,\bar{\m},\a,\bar{\a}\}.
\eea
The operator $\d$ is of total degree 0 and obeys the following
relations
\be
d=-[s,\d]~~~~,~~~~[d,\d]=0.
\ee
The operator $\d$ can be used to solve the descent equations
\equ{des}. As it
has been shown in \cite{st,mss,tv} these solutions can be
obtained from the
equation
\be
\label{solutia2}
(s+d)\o_0(c+dz+\m d\bar{z} ,\bar{c}+d\bar{z}+\bar{\m} d z ,
\e -\a d\bar{z},
\bar{\e} -\bar{\a} d z, X,\l,\bar{\l}, F,\m,\bar{\m},\a,
\bar{\a})=0
\ee
where $\o_0$ is a solution of the equation \equ{basic}, by
projecting out
different terms with a given ghost and space-time degree.

The main purpose of our paper is to solve the
equation ~\equ{basic} in the algebra of all local polynomials of
the
fields $\AA$. A basis for this algebra can be chosen to consist of
\be
\left\{ \6^p\bar{\6}^q\Psi,\/\6^p\bar{\6}^qc\/,
\/\6^p\bar{\6}^q\bar{c}\/,
\/\6^p\bar{\6}^q\e\/,\/\6^p\bar{\6}^q\bar{\e} \right\},
\label{baza}
\ee
where $\Psi=\{X\/,\/\l\/,\/\bar{\l}\/,\/F\/,\/\m\/,\/\bar{\m}\/,
\/,\a\/,
\/\bar{\a}\}$. However, the BRST transformations of this basis is
quite complicated and it contains many terms which can be
eliminated in H(s). In the next section we shall eliminate
a part
of the basis ~(\ref{baza}) and in the following one we shall
introduce
a new basis, where the action of the BRST differential $s$ is
very simple.


\section{Contractive algebra and Sullivan Theorem}
\setcounter{equation}{0}
The calculation of the BRST cohomology ,i.e. the solutions of
the equations
\be
s\o_0 = 0 \label{fund}
\ee
can be considerably simplified if we
organize the  algebra $\AA$ as a free differentiable algebra
and use a very
strong and efficient theorem due to D.Sullivan \cite{sull}.
A free
differential algebra is an algebra generated by a basis endowed
with a
differential. Sullivan's tells us that
the most general free differential algebra $\AA$ is a tensor
product
of a contractible algebra $\CC$ and a minimal  one $\MM$ .
A minimal differential algebra $\MM$ with the differential $s$
is an algebra
for which $\MM\subseteq\MM^+\MM^+$ where $\MM^+$ is the part of
$\MM$ in
positive degree, i.e. $\MM=C\oplus\MM^+$ and a contractive
differential
algebra $\CC$ is one isomorphic to the tensor product of those of
the form $\L(x,sx)$.

On the other hand, due to K\"{u}neth theorem the cohomology of
$\AA$
is given by the cohomology of its minimal part $\MM$ and we can
say that
the contractible subalgebra $\CC$ can be neglected in the
calculation
of the cohomology H(s). In \cite{af,cfgp,caf} was given a
general iterative
construction of the minimal differential algebra of a given
free differential
algebra $\AA$ and this construction can be applied in our case.
However, for our differential algebra the construction of $\CC$
and
$\MM$ is straightforward and we do not have to rely on the
general
construction. In fact it is easy to see from \equ{brst3}
and \equ{brst4}
that the generators of \equ{baza} of $\AA$ which have the form
\be
\{ \6^p\bar{\6}^q (\bar{\6}c)~~,~~\6^p\bar{\6}^q (\6\bar{c})~~,~~
\6^p\bar{\6}^q (\bar{\6}\e)~~,~~\6^p\bar{\6}^q(\6\bar{\e})\} \ee
with $p,q=0,1,\cdots$ can be replaced by
\be
\{ \6^p\bar{\6}^q\PH~,~s( \6^p\bar{\6}^q\PH)~,~\6^p c~,~\bar{\6}^p
\bar{c}
~,~\6^p \e~,~\bar{\6}^p\bar{\e}\} \ee

with $\PH=\{ \m~,~\bar{\m}~,~\a~,~\bar{\a}\}$ and the algebra
$\AA$ could be
generated by
\be
\{  \6^p\bar{\6}^q\PH~,~s( \6^p\bar{\6}^q \PH)~,~
\6^p\bar{\6}^q(\Psi)~,~
\6^p c~,~\bar{\6}\bar{c}~,~\6^p \e~,~\bar{\6}^p \bar{\e}\}
\label{newbasis}\ee
where $\Psi=(X~,~\l~,~\bar{\l}~,~F)$. Now the Sullivan
decomposition
can be easily obtained from \equ{newbasis} since the
contractible algebra
$\CC$ is generated by
\be
\label{contactible}
\{  \6^p\bar{\6}^q\PH~~~,~~~s( \6^p\bar{\6}^q\PH) \}
\ee
and the minimal subalgebra $\MM$ might be generated by the
elements
\be
\{  \6^p\bar{\6}^q\Psi~~,~~\6^p c~~,~~\bar{\6}\bar{c}~~,~~\6^p
\e~~,~~
\bar{\6}\bar{\e}\}. \label{minimal}
\ee

However, the basis \equ{minimal} might not generate a minimal
differential
algebra since the derivatives of $\Psi$ do not have simple BRST
transformations and it is difficult to see whether the condition
$\MM\subseteq\MM^+\cdot \MM^+$ is satisfied or not.
The situation can be considerable improved if one introduces a
new basis.
To define this new basis we introduce four linear even operators,
which will play a crucial role in our considerations:
\bea
\D=\{s~,~\f{\6}{\6 c}\}~~~,~~~\bar{\D}=\{s~,~\f{\6}{\6\bar{c}}\}
\label{basis}\\
\D_0=\{s~,~\f{\6}{\6(\6 c)}\}~~~,~~~\bar{\D}_0
=\{s~,~\f{\6}{\6\bar{\6}\bar{c}}\}\label{weight}
\eea
These operators are, in fact, even differentials since they satisfy
Leibniz
rule $$
D(a b)=(D a)b+a (D b) $$
where $D$ is one of the operators just defined.

The operators \equ{basis} could replace the usual derivatives in
the new basis defined by
\be
\PH^{p,q}=\D^p\bar{\D}^q\PH ~~~\mbox{where}~~\PH=\{X,\l,\bar{\l},F\}
\label{newbasisd}
\ee
and
\bea
c^p=\f{1}{(p+1)!}\D^{p+1}c=\f{1}{(p+1)!}\6^{p+1}c &,&
\bar{c}^p=\f{1}{(p+1)!} \bar{\D}^{p+1}\bar{c}
=\f{1}{(p+1)!}\bar{\6}^{p+1}
\bar{c} \non
\e^{p+\f{1}{2}}=\f{1}{2}\f{1}{(p+1)!}\D^{p+1}\e
=\f{1}{(p+1)!}\6^{p+1}\e &,&
\bar{\e}^{p+\f{1}{2}}=\f{1}{2}\f{1}{(p+1)!}\bar{\D}^{p+1}\bar{\e}^p
=\f{1}{(p+1)!}\bar{\6}^{p+1}
\bar{\e} .\label{newbasist}
\eea
where $p=-1,0,+1,\cdots $.

The action of the BRST differential $s$ on the new basis
can be easily obtained
if we use the commutation relations
\be
[\D~,~s]=0~~~~~,~~~~[\bar{\D}~,~s]=0 \label{comm100} \ee
Thus eqs. \equ{brst1} and \equ{comm100} yield
\be
s\PH^{p,q}=\sum_{k=-1}(c^k L_k+\bar{c}^k\bar{L}_k +\e^{k+\f{1}{2}}
G_{k+\f{1}{2}} +\bar{\e}^{k+\f1{2}}\bar{G}_{k+\f1{2}})
\PH^{p,q}  \label{brst100}
\ee
where
\bea
L_k \PH^{p,q} & = & A^p_k(a_{\PH})\PH^{p-k,q}~~~~,
~~~~\bar{L}_k\PH^{p,q}\hspace{0.75cm}=~~A^q_k(b_{\PH}) \PH^{p,q-k}
\non
G_{k+\f1{2}}\PH^{p,q} & = & A_{k+1}^p\PH_{\e}^{p-k-1,q}~~~~
,~~~~\bar{G}_{k+\f1{2}}\PH^{p,q}~~=~~A_{k+1}^q
\PH_{\e}^{p,q-k-1} ,\label{g}
\eea
with
\be
A^{p}_{k}(a) =\f{p!}{(p-k)!}[p-k+a(k+1)]~~,~~A^p_k=\f{p!}{(p-k)!}.
\label{A}
\ee
In eqs.\equ{g} the fields $ \PH_{\e}^{p,q} $ have the form given
by eq.
\equ{newbasisd}, i.e.
\be
\PH_{\e}^{p,q}= \D^p \bar{\D}^q \PH_{\e}
\ee
and are given by eqs.\equ{e1}, \equ{e2}, \equ{e3}, \equ{e4}.

The linear operators $ \{ L_k, \bar{L}_k, G_{k+\f1{2}},
\bar{G}_{k+\f1{2}} \} $
represent on $ \PH^{p,q} $ the super-Virasoro algebra since they obey on
$ \PH^{p,q} $ the (anti)commutation relations:

\be
[ L_m , L_n ]  =  (m-n)L_{m+n} ~~,~~
 \{ G_a , G_b \}  =  2L_{a+b} ~~,~~
[ L_n , G_a ]  =  (\f{n}{2}-a)G_{a+b}
\ee

where $m, n = -1, 0, 1, \cdots $ and $a, b = -\f1{2}, \f1{2}, \cdots $ and
the same relations with $L_m $ and $G_a $ replaced by
$\bar{L}_m $ and $\bar{G}_a $. In addition
\bea
[L_m , \bar{L}_n ] =[L_m,\bar{G}_a ] & = & [G_a , \bar{L}_m] =0 \\
\{ G_a , \bar{G}_b \} & = &0
\eea
Thus the generators of the BRST transformations on $\PH^{p,q}$
form two copies
of the super-Virasoro algebra without central charge. All these
operators
can be defined on the whole algebra by fields $ \PH^{p,q} $ as
even or odd derivatives.
In order to do that we just notice that from \equ{g} one can write
\be
L_k = \{ s,\f{\6}{\6 c^k} \}~~,~~G_{k+\f1{2}}=[\f{\6}{\6
\e^{k+\f{1}{2}} },s]
\ee
and by taking into account that $s$ and $\f{\6}{\6 c^k } $ are
 odd derivatives
and $\f{\6}{\6 \e^{k+\f1{2}}}$ are even one can extend $L_k $ and
$G_{k+\f1{2}}$ as even and odd derivatives, respectively.

The BRST transformations
of the ghosts basis
$ \{ c^{n}~,~{\bar{c}}^n~,~\e^{n+\f1{2}}~,~{\bar{\e}}^{n +\f1{2}} \} $
$(n=-1,0,1, \cdots)$ can easily be written as:

\bea \label{brst101}
s c^n & = & \f1{2} f^n_{pq} c^q c^p -\f1{2}\bar{f}^n_{ab}\e^a\e^b\non
s \e^a & = & -\tilde{f}^a_{mb}c^m \e^b
\eea
where
\be
f^n_{pq}=(p-q)\d^n_{p+q}~~,~~ \bar{f}^n _{ab} = 2\d^n_{a+b}~~,
~~\tilde{f}^a_{mb} = (\f{m}{2} - b)\d^a_{m+b}
\ee
and $p,q = -1, 0, \cdots, a=-\f1{2}, \f1{2},\cdots$.
The new basis \equ{newbasisd} and \equ{newbasist} have been introduced
by Brandt, Troost and Van Proeyen \cite{btv1} for 2D-gravity and was
used
in \cite{tv} for Beltrami parametrization.

The main result of all these considerations is the fact that the
subalgebra
$\MM$ generated by
$$
\{\PH^{pq}~~,~~c^p~~,~~\bar{c}^p~~,~~\e^p~~,~~\bar{\e}^p \}
$$
is a minimal algebra and its BRST cohomology coincide with
H(s). Therefore,
at this stage to calculate the BRST cohomology H(s) we simply
have to
consider only the minimal subalgebra $\MM$ together with its BRST
transformations \equ{brst100} and \equ{brst101}.

In fact one can reduce the number of the candidates for the members of H(s)
if we use the operators $\D_0~,~\bar{\D}_0$ introduced in \equ{weight}.
They have the remarkable property that all the elements of the basis of
$\MM$ are their eigenfunctions:
\bea
\D_0\PH^{pq} & = & (a_{\PH}+p)\PH^{pq} \hspace{0.5cm},
\hspace{0.5cm}\bar{\D}_0\PH^{pq}=(b_{\PH}+q)\PH^{pq}\non
\D_0 c^p & = & p c^p          \hspace{2cm},
\hspace{0.5cm}\bar{\D}_0 c^p=0                     \non
\D_0\bar{c}^p & = & 0         \hspace{2.2cm},
\hspace{0.5cm}\bar{\D}_0\bar{c}^p=p \bar{c}^p       \non
\D_0\e^a & = & a\e^a          \hspace{2cm},
\hspace{0.5cm}\bar{\D}_0\e^a=0                  \non
\D_0\bar{\e}^a & =& 0         \hspace{2.2cm},
\hspace{0.5cm}\bar{\D}_0\bar{e}^a=a\e^a. \non
\eea
where $p,q=-1,0, 1,\cdots , a=-\f1{2},\f1{2},\f{3}{2},\cdots$.
Therefore all monomials from $\MM$ are eigenvalues of $\D_0$
and $\bar{\D}_0$
and their eigenvalues are, by definition, {\em the total weight}.
Due to
the fact that $\D_0$ and $\bar{\D}_0 $ are the anticommutators
of $s$ with
something it is easy to convince ourselves that a solution of
the equation
$s\o_0=0$ with a total weight different form (0,0) is $s$-exact,
i.e.
if
$$
s\o_0=0~~~,~~~\D_0\o_0=p\o_0~~(p=1,2,\cdots)
$$
then
$$
\o_0 = s(\f1{p}\f{\6 \o_0}{\6 (\6 c)}).
$$

\section{The new basis and BRST cohomology}
\setcounter{equation}{0}

There is an essential difference between the Beltrami and
super-Beltrami
parameterizations when one tries to calculate the BRST
cohomology. For the
2D gravity Brandt, Troost and Van Proeyen \cite{btv2}
(see also \cite{btv1,btv,tv})
have shown that the basis of the BRST cohomology H(s), contains
a finite
number of terms with  very simple forms. For the
super-Beltrami
parametrization this is not the case due to the
presence of the bosonic ghosts
$\e$ and $\bar{\e}$. In this case, even if we impose the
condition of the
weigh to be $(0,0)$, we obtain a basis for the BRST cohomology
group H(s)
with an infinite number of terms. Besides, the form of these
terms is much more
involved in comparison with the Beltrami parametrization.
Therefore, we are
not going to give a complete discussion but rather we shall
analyze only the
general solution of \equ{fund} for ghost number two and we give
some particular
solutions for ghost number three. They are in fact the
supersymmetric extensions
of the BRST cohomology elements
for the non-supersymmetric case, that is, for the Beltrami
parametrization
\cite{btv,tv}.

For ghost number two we consider the general solution of \equ{fund}
of the form
\be
\o_0=c\bar{c}\P_1 +\e\bar{c}\P_2+\bar{\e} c\P_3
+\e\bar{e}\P_4+\e c\P_5 +\e\bar{c}\P_6
\label{sup}
\ee
the total weight of $\P_j , (j=1,2,3,4)$ must have the values
$$
w(\P_1)=(1,1)~,~w(\P_2)=(\f1{2},1)~,~w(\P_3)=(1,\f1{2})~,
~w(\P_4)=(\f1{2},\f1{2})
$$
and they are polynomials in the classical fields and their
derivatives,
not containing any ghost. Now we want to impose the
conditions that
${\o}_0$ is a solution of eq.\equ{fund}. Taking into
consideration the
BRST transformations of all ghosts \equ{brst4} and
\equ{brst5} and
for the classical fields \equ{brst100} the equation
\equ{fund} yields
\be
\label{cond}
L_k\P_A=\bar{L}_k\P_A=G_a\P_A=\bar{G}_a\P_A=0
\ee
with $k=1,2,\dots, a=\f{3}{2},\f{5}{2},\cdots$ and $A=1,2,3,4$,
\bea
\label{weight1}
L_0\P_1 & = & \P_1~~~,~~~\bar{L}_0\P_1=\P_1 \non
L_0\P_2 & = & \f1{2}\P_2~~~,~~~\bar{L}_0\P_2=\P_2 \non
L_0\P_3 & = & \P_3~~~,~~~\bar{L}_0\P_3=\f1{2}\P_3 \non
L_0\P_4 & = & \f1{2}\P_4~~~,~~~\bar{L}_0\P_4=-\f1{2}\P_4
\eea
and
\bea
\label{weight2}
\P_1 & = & -G_{-\f1{2}}\P_2~~,~~\P_1~~=
~~-\bar{G}_{-\f1{2}}\P_3\non
\P_2 & = & -G_{\f{1}{2}}\P_1~~,~~\P_3~~=
~~\bar{G}_{\f1{2}}\P_1~~,\non
\P_4 & = & -G_{\f1{2}}~~,~~
\P_3~~,~~\P_4~~=~~\bar{G}_{\f1{2}}\P_2\non
\P_2 & = & \bar{G}_{-\f1{2}}\P_4 ~~,~~\P_3~~=
~~G_{-\f1{2}}\P_4\non
G_{-\f1{2}}\P_1 & + & L_{-1}\P_2 =0,
\bar{G}_{-\f1{2}}\P_1 + \bar{L}_{-1}\P_3 =0
\eea
The equations \equ{weight} show that $\P_A~(A=1,2,3,4)$ have
the correct
weight. The equations \equ{weight2} indicate the fact that all
$\P_A~(=1,2,3,4)$
can not depend on $\Phi^{p,q}$ with $p\geq 1$ or $q\geq 1$.

The equations \equ{weight2} impose some conditions on $\P_1$
and give
the possibilities to calculate $\P_2~,~\P_3$ and $\P_4$ as far
as we
know $\P_1$. In addition $\P_1$ must contain the terms  from
Beltrami
parametrization \cite{btv,bv} i.e. it has to be of the form
\be
\P_1=X^{1,0} X^{0,1}+\cdots
\label{ansaz}
\ee
With this "boundary condition" equations \equ{weight2} yield
\bea
\label{final}
\P_1 & = & X^{1,0}X^{0,1}-\l\l^{0,1}-\bar{\l}\bar{\l}^{1,0}-F^2
\non
\P_2 & = & -\l X^{0,1}-iF\bar{\l}   \non
\P_3 & = & X^{1,0}\bar{\l}-\l\bar{\l}   \non
\P_4 & = & \l\bar{\l} \non
\eea
The supersymmetric invariant action can be calculated to be the
term
$\o_2$ in the local descent equation \equ{des} as long as
$\o_0$ is given by
Eqs. \equ{sup} and \equ{final}. It can be, in fact, found out
by projecting
out in equation \equ{solutia2} the terms with ghost zero.
Therefore we
eventually get
\bea
-2i S_{inv}&=&\int dz\wedge d\bar{z}
\{ (1-\m\bar{\m}\P_1 +
\f1{2}\bar{\m}\a\P_2 -
\f1{2}\m\bar{\a}\P_3 +\f1{4}\P_4) \} \non
&=&\int dz\wedge d\bar{z}
\{\f1{1-\m\bar{\m}}
[(\6-\bar{\m}\bar{\6})X(\bar{\6}-
\m\6)X \non
&-&(\a\l)(\6-\bar{\m}\bar{\6})X -
(\bar{\a}\bar{\l})(\bar{\6}-\m\6)X +
\f1{2}(\a\l)(\bar{\a}\bar{\l})]\non
&-&\l(\bar{\6}-\m\6)\l -\bar{\l}(\6 -\bar{\m}
\bar{\6})\bar{\l}-(1-\m\bar{\m})F^2\}
\eea
The action obtained by us has the same form with the one
obtained by Delduc
and Gieres \cite{dg} .

At this point it is necessary to make the following remarks.
First the solution
of the equations \equ{weight1} and \equ{weight2} is not
unique and in
addition to the solution just given there is another one
with $\P_4$
given by
\be
\P_4=F.
\ee
In this case the solution takes the form:
\be
\o_0=-i c\bar{c}+i\e\bar{c} \l^{0.1}-i \bar{\e}c\bar{\l}^{1,0}
+\e\bar{\e} F.
\ee
However this solution could not be used for building of an
invariant
action since it is linear in fields.
Second, even in the case of the first solution one has a
freedom in choosing
the "boundary conditions". As a matter of fact Brand, Troost
and
Van Proeyen \cite{btv2} showed that in the two dimensional gravity
the most general
solution of the equation \equ{fund} has the form
\be
\o_0=c\bar{c}X^{1,0}X^{0,1} f(X)
\ee
with $f(X)$ an arbitrary function of $X$. We could obtain
the supersymmetric
extension of this solution if one starts with the following
expression for
$\P_4$
\be
\P_4=\l\bar{\l} f(X).
\ee
In this case the solution of equations \equ{weight1} and
\equ{weight2}
have the form
\bea
\label{final1}
\P_1 & = & (X^{1,0}X^{0,1}-\l\l^{0,1}-\bar{\l}\bar{\l}^{1,0}
-F^2)f(X)+iF\l
\bar{\l}f'(X)  \non
\P_2 & = &( -\l X^{0,1}-iF\bar{\l})f(X)   \non
\P_3 & = &( X^{1,0}\bar{\l}-\l\bar{\l})f(X)   \non
\P_4 & = & \l\bar{\l}f(X) \non
\eea
where the prime means the derivative. In this case the
supersymmetric action
contains the arbitrary function $f(X)$ and it has the form
\bea
-2i S_{inv}&=&\int dz\wedge d\bar{z} \{ (1-\m\bar{\m}\P_1 +
\f1{2}\bar{\m}\a\P_2 -\f1{2}\m\bar{\a}\P_3 +\f1{4}\P_4 \}\non
&=& \int dz\wedge d\bar{z} \{\f1{1-\m\bar{\m}}[(\6-
\bar{\m}\bar{\6})X(\bar{\6}-
\m\6)X\non
&-&(\a\l)(\6-\bar{\m}\bar{\6})X -(\bar{\a}\bar{\l})
(\bar{\6}-\m\6)X +
\f1{2}(\a\l)(\bar{\a}\bar{\l})]\non
&-&\l(\bar{\6}-\m\6)\l -\bar{\l}(\6 -\bar{\m}
\bar{\6})\bar{\l}-(1-\m\bar{\m})F^2\}f(X)\non
&+&i\int dz\wedge d\bar{z} F\l\bar{\l}f'(X).
\label{mostgeneral}
\eea
with $f'(X)$ the derivative of $f(X)$. The action
\equ{mostgeneral} is the
most general supersymmetric invariant action which depends on
the matter
chiral superfield with the components $X,~\l,~\bar{\l},~F$ and
2D supergravity
with the components $\m,~\bar{\m},~\a,~\bar{\a}$. It is
interesting to
notice that only the field $X$ can occur in the general
dependence form.
Our general result can be easily extended for the matter
field described by
a set of chiral superfields, extension which will be
given elsewhere \cite{tv2}.

Now we offer the results of our analyses for ghost number three.
This case is very important since it leads to the possible
anomalies in the
super-Beltrami parametrization. The BRST
cohomology for Beltrami parametrization was given in
\cite{btv} (see also \cite{tv})
and it contains only four terms
\be
c^{-1}c^{0}c^1~,~\bar{c}^{-1}\bar{c}^{0}\bar{c}^1~,
~c^{-1}\bar{c}^{-1}c^0
X^{1,0}X^{0,1}f(X)~,~c^{-1}\bar{c}^{-1}\bar{c}^0X^{1,0}X^{0,1}f(X)
\label{beltrami1}
\ee
where $f(X)$ is an arbitrary function.

We find out the supersymmetric extensions of
all these solutions.
The element $c^{-1}c^{0}c^1$ does contain only ghosts and
its supersymmetric
extension must have the form
\be
\o_0= c^{-1}c^{0}c^1 +ac^{-1}\e^{\f1{2}}\e^{\f1{2}}+
bc^0\e^{-\f1{2}}\e^{\f1{2}}
\ee
being the unique combination of ghost monomials with
the weight (0,0).
Here $a$ and $b$ are two numerical constants,
which can be determined from the
equation \equ{fund}. If we take into consideration the
BRST transformation
of $c^{-1}$ and $\e^{-\f1{2}}$ and their derivatives
\equ{brst4} , \equ{brst5} one can
readily
find out that $a=-2$ and $b=1$, that is the supersymmetric
extension of
$c^{-1}c^{0}c^1$ is
\be
\o_0^1=c^{-1}c^{0}c^1-2c^{-1}\e^{\f1{2}}\e^{\f1{2}}+
c^0\e^{-\f1{2}}
\e^{\f1{2}}.
\ee
Analogously, the supersymmetric extension of
$\bar{c}^{-1}\bar{c}^{0}\bar{c}^1$
turns out to be
\be
\o_0^2=\bar{c}^{-1}\bar{c}^{0}\bar{c}^1-
2\bar{c}^{-1}\bar{e}^{\f1{2}}
\bar{\e}^{\f1{2}}+\bar{c}^0\bar{\e}^{-\f1{2}}\bar{\e}^{\f1{2}}.
\ee
The corresponding analogy can be obtained from
$\o_0^1$ and $\o_0^2$ by applying the formula \equ{delta} and
projecting out the ghost number one. In this way one gets
\bea
\AA_1 & = & 4 \int dz\wedge d\bar{z} (c\6^3 \m +\e \6^2 \alpha)
\non
\AA_2 & = & 4\int dz\wedge d\bar{z} (\bar{c}\bar{\6}^3\bar{\m}+
         \bar{\e}\bar{\6}^2\bar{\alpha})
\eea
The supersymmetric extension of
$c^{-1}\bar{c}^{-1}c^0 X^{1,0}X^{0,1} f(x)$
may be calculated by the same method used previously
for ghost number two.
We start with a solution of eq.\equ{fund} of the form
\bea
\o_0^3 & = & c^{-1}\bar{c}^{-1}c^0 \P_1 \non
       & + & \e^{-\f1{2}}\bar{c}^{-1}c^0 \P_2
       +c^{-1}\bar{\e}^{-\f1{2}}c^0 \P_3 +
c^{-1}\bar{c}\e^{\f1{2}}\P_4 \non
       & + & \e^{-\f1{2}}\bar{\e}^{-\f1{2}}c^0 \P_5 +
       \e^{\f1{2}}\bar{c}^{-1}\e^{\f1{2}}\P_6
       + c^{-1}\bar{e}^{-\f1{2}}\e^{\f1{2}}\P_7
       +\e^{-\f1{2}}\bar{\e}^{-\f1{2}}\e^{\f1{2}}\P_8
\label{czero}
\eea
The total weight of $\P_A~~(A=1,2,3,4,5,6,7,8)$ have the values
\bea
w(\P_1)=(1,1) &,& w(\P_2)=(\f1{2},1) ~~,~~w(\P_3)=(1,\f1{2}) \non
w(\P_4)=(\f1{2},1)&,& w(\P_5)=(\f1{2},\f1{2}) ~~,~~w(P_6)=
(\f1{2},\f1{2})
\eea
The equation \equ{fund} yields a set of equations very
similar
with the ones used in the ghost  case. We are not going
to write them down
(see \cite{tv2} for details) but rather we will give only the
final
result. Therefore the solution of equation \equ{fund} of the
form  \equ{czero}
is
\bea
\o_0^3 & = & [ c^{-1} \bar{c}^{-1}c^{0} (X^{1,0}X^{0,1}-F^2
         + \bar{\l}^{1,0}\bar{\l} -\l\l^{0,1}) \non
       & + & (2\e^{\f1{2}}c^{-1}\bar{c}^{-1} +
         \e^{-\f1{2}}c^0\bar{c}^{\f1{2}}) (iF\l +
          X^{0,1}\bar{\l})\non
       & - & \bar{\e}^{-\f1{2}}c^{-1}c^0(-i\l F+
          X^{1,0}\bar{\l})\non
       & + &\e^{-\f1{2}}\bar{\e}^{-\f1{2}}c^0\f1{2}\l\bar{\l} -
          2\bar{\e}^{-\f1{2}}\e^{\f1{2}}c^{-1}\l\bar{l}]f(X)\non
       & - &ic^{-1}\bar{c}^{-1}c^{0}F(\l\bar{\l})f^{'}(X)
\eea
In a similar manner one can obtain the supersymmetric extension
of the
$c^{-1}\bar{c}^{-1}\bar{c}^0 X^{1,0}X^{0,1} f(X)$.It has the form
\bea
\o_0^4 & = & c^{-1}\bar{c}^{-1}\bar{c}^0 \P_1 \non
       & + & \e^{-\f1{2}}\bar{c}^{-1}\bar{c}^0 \P_2
       +c^{-1}\bar{\e}^{-\f1{2}}\bar{c}^0 \P_3 +
       c^{-1}\bar{c}\bar{\e}^{\f1{2}}\P_4 \non
       & + & \e^{-\f1{2}}\bar{\e}^{-\f1{2}}\bar{c}^0 \P_5 +
       \e^{\f1{2}}\bar{c}^{-1}\bar{\e}^{\f1{2}}\P_6
       + c^{-1}\bar{e}^{-\f1{2}}\bar{\e}^{\f1{2}}\P_7
       +\e^{-\f1{2}}\bar{\e}^{-\f1{2}}\bar{\e}^{\f1{2}}\P_8
\label{bczero}
\eea
with the corresponding weights for the polynomials $\P_i$.
Performing the same algebraic calculations, one
gets
\bea
{\o}_0^4 & = & [ c^{-1} \bar{c}^{-1}\bar{c}^{0} (X^{1,0}X^{0,1}- F^2
               + \bar{\l}^{1,0}\bar{\l} -\l\l^{0,1}) \non
         & - & (\bar{\e}^{-\f1{2}}c^{-1}\bar{c}^{0}
           -  2\bar{\e}^{\f1{2}}\bar{c}^{-1}c^{-1})(\bar{\l}X^{1,0}
           -   i\l F) \non
         & - & \e^{-\f1{2}}c^{-1}\bar{c}^{0}(\l X^{0,1}+
           iF\bar{\l})+(\e^{-\f1{2}}\bar{\e}^{-\f1{2}}\bar{c}^{0}\l\bar{\l}
             -2\e^{\f1{2}}\bar{\e}^{\f1{2}}\bar{c}^{-1}\l\bar{\l}]f(X) \non
         & - &ic^{-1}\bar{c}^{-1}\bar{c}^{0}f^{'}(X)F(\l\bar{\l})
\eea
Using again \equ{solutia2} and projecting out the terms with
ghost number 1 we can find the possible anomalies. They have
the form
\bea
\AA_3 & = & \int dz\wedge d\bar{z} \{ [\6 \m (\bar{\m}c -\bar{c})
        + \6 c(1-\m \bar{\m})][(X^{1,0}X^{0,1} -F^2 +
\bar{\l}^{1,0}\bar{\l}
        - \l\l^{0,1})](f(X)-if^{'}(X)F\l\bar{\l}) \non
      & + & \{ [\bar{\m}(\e\6\m +\alpha\6 c +
2\6\alpha(c-\m \bar{c})
        + 2\6\e(1-\m\bar{\m})](\l X^{0,1} +iF\bar{\l}) \non
      & + & [\6\m (\bar{\alpha}c+\bar{\e})
        - \m\bar{\alpha}\6 c](-X^{1,0}\bar{\l}+i\l F) \non
      & + & [-\bar{\alpha}(\alpha\6 c + \e\6\m ) +
2 \m(\bar{\e}\6\alpha
        + \6\e\bar{\alpha})]\l\bar{\l} \}f(X)\}
\eea
and
\bea
\AA_4 & = & \int dz\wedge d\bar{z} \{ [\6 \bar{\m}
          (\m\bar{c} -c)
        + \6 \bar{c}(1-\m \bar{\m})][(X^{1,0}X^{0,1} -
           F^2 +\bar{\l}^{1,0}\bar{\l}
        - \l\l^{0,1})](f(X)-if^{'}(X)F\l\bar{\l}) \non
      & + & \{ [\m(\bar{\e}\6\bar{\m} +\bar{\alpha}\6 \bar{c}
        + 2\6\bar{\alpha}(\bar{c}-\bar{\m} c)
        + 2\6\bar{\e}(1-\m\bar{\m})](\bar{\l} X^{1,0} -iF\l) \non
      & + & [\6\bar{\m} (\alpha \bar{c}+\e)
        - \bar{\m}\alpha\6\bar{c}](X^{0,1}\l+i\bar{\l} F) \non
      & + & [\alpha(\bar{\alpha}\6\bar{c} + \bar{\e}\6\bar{\m} )
        +2\bar{\m}(\e\6\bar{\alpha}
        +\6\bar{\e}\alpha)]\l\bar{\l} \}f(X) \}
\eea
As in case of ghost two we could have started the investigation
of
equations for $\P_a $ with $\P_7 = F$. However, it seems to us
that
all these anomalies cannot be associated with true anomalies.
Indeed,
in the classical action \equ{mostgeneral} the terms of
self-interactions
are absent and in spite of the fact that these anomalies are
cosmologically
non-trivial, the numeric coefficients of the corresponding
Feynman diagrams vanish.

\section{Conclusions}
\setcounter{equation}{0}

We studied a part of the BRST cohomology for super-string in
super-Beltrami parametrization with a given field content
(super-Beltrami
parametrization and Chiral matter superfield) and gave gauge
invariances
(superdiffeomorphism invariance). We have given a partial answer
for the cohomology not only on local functions but also on local
functionals,
the latter being obtained from the former by the descent
equations or equivalently
by using the operator $\d$ introduced by Sorella. In this way we
have
obtained the most general classical action describing the
superstrings
in super-Beltrami parametrization given by \cite{dg}, (see also
\cite{mss} ). Also we have found out the
candidate anomalies, which are supersymmetric extension of
the corresponding
non-supersymmetric ones given by Brandt, Troost and Van Proeyen
\cite{btv2}( see
also \cite{s} and \cite{tv} ). They are of two types. The first
type do not depend
on the matter fields and their combination provides the
supersymmetric extension
of the Weyl anomaly \cite{btv}. The second type anomalies
depend on the matter
fields and might also depend on the arbitrary functions of
the scalar part $X$
of the matter field.
In the present paper the antifields and the Batalin-Vilkovisky
method of
quantization have not been taken into consideration. In this
way we are aware
that we have lost a lot of information concerning the structure
and the
symmetries of the model. We shall study the BV-method in
the forthcoming
publication \cite{tv2}.

Finally, we want to notice that all these calculations could
be done in
the superfield formulation of the theory. In this case we
work only with
superfields (super-Beltrami, superghosts and supermatter)
and it is possible
to introduce the Sorella's operator $\d$ even for this case.
We intend to
give a superfield formulation of the BRST cohomology in a future
publication.


\begin{thebibliography}{99}
\bibitem{gat}S. J. Gates, Jr. and H. Nishino,\cqg {3}
{86}{391}; M. Rocek, P. van
Nieuwenhuizen and S.C. Zhang,\annp{172}{86}{348};
\bibitem{br}R.Brooks, S. J. Gates, Jr and F. Muhammad,\np{B268}
{86}{599}; M. T. Grisaru, and Rui-Ming Xu \pl{205B}{88}{486}
\bibitem{gir} S. J. Gates,Jr. and F. Gieres,\np{B320}{89}{310}
\bibitem{gr} M. T.Grisaru and Marcia E. Wehlau,\np{B453}{95}{489}
\bibitem{ht} M.Henneaux and C.Teitelboim ,{\em Quantization of
Gauge
Systems}, Princeton University Press,1992
\bibitem{wz1}J.Wess and B.Zumino, \pl{B37}{71}{95}
\bibitem{bv}I.A.Batalin and G. A. Vilkovisky, \pl{B102}{81}
{27}\pr{D28}{83}{2567}
\bibitem{s}S.P.Sorella, \cmp{157}{93}{231}
\bibitem{wss}M.Werneck de Oliveira, M.Schweda and S.P.Sorella,
 \pl{B315}{93}{93}
\bibitem{bdk} F.Brandt, Dragon and M.Kreutzer,\np{B332}{90}{224}
\bibitem{dg}F.Delduc and F.Gieres, \cqg{7}{90}{1907}
\bibitem{bbg1} L.Baulieu, M.Belolon and R.Grimm, \pl{B198}{87}{347}
\bibitem{bbg2} L.Baulieu, M.Belolon and R.Grimm, \np{B321}{89}{697}
\bibitem{g} R.Grimm, \annp{200}{90}{49}
\bibitem{bss}A.Boresch, M.Schweda and S.P.Sorella, \pl{B328}{94}{36}
\bibitem{sull} D. Sullivan {\em Infinitesimal computations in
topology},
Bulletin de l'Institute des Hautes Etudes
 Scientifiques, Publication Mathematique nr. 47 (1977)
\bibitem{st}S.P.Sorella and L.Tataru, \pl {B324}{94}{351}
\bibitem{mss}O.Moritsch, M.Schweda and S.P.Sorella, \cqg{11}{94}{1225}
\bibitem{tv}L.Tataru and I.V.Vancea \ijmp {\bf 11} 2 (1996) 375-393
\bibitem{af} R.D'Auria, P.Fr\'{e}, \np{B201}{82}{101}
\bibitem{cfgp} L.Castellani, P.Fr\'{e}, G.Gianini, K.Pilch,
P.van
Nieuweinhuisen, \annp{146}{83}{35}
\bibitem{caf}L.Castellani, R.D'Auria, P.Fr\'{e}, {\em
Supersymmetry and
Superstring, 1983, Proceeding of the XIXth Winter School and
Workshop
of Theoretical Physics}, Word Scientific, (1983)
\bibitem{btv1} F. Brandt, W. Troost and A. Van Proeyen ,
NIKHEF-H 94-16, KUL-TF-94/17, hep-th/9407061, to appear in
the proceedings of the {\em Geometry of Constrained Dynamical
Systems} workshop, Isaac Newton Institute for Mathematical
Sciences, Cambridge, June 15-18, 1994;  W. Troost and A. Van
Proeyen, KUL-TF-94/94, hep-th/9410162
\bibitem{bl1}G.Bandelloni and S. Lazarini, \jmp{34}{93}{5413}
\bibitem{bl2}G.Bandelloni and S. Lazarini, {\em Diffeomarphism
cohomology in Beltrami parametrization II: the 1-forms},
PACS 11.10.Gh/03.70- University of Marseille reprint ( to be
published in
{\em Journal of Mathematical Physics})
\bibitem{witten}E.Witten, \np{B373}{92}{187}
\bibitem{wz2}E.Witten and B.Zwiebach, \np{B377}{92}{55}
\bibitem{btv}P.A.Blaga, L.Tataru and I.V.Vancea,
{\em Roumanian Journal of Physics}{\bf 40} 10 457 (1995)
\bibitem{btv2} F.Brandt, W.Troost and A.Van Proeyen,
{\em The BRST-antibracket
cohomology of 2D gravity conformally coupled to scalr matter},
University of Leuven preprint, KUL-TF-95/17, hep-th/9509035
\bibitem{tv2} L.Tataru and I.V.Vancea, BRST-BV cohomology of the
superstring in the super-Beltrami parametrization
(paper in preparation).
\end{thebibliography}
\end{document}